\pdfoutput=1

\documentclass[11pt,reqno]{article}
\usepackage{jheppub}
\usepackage{epsfig}
\usepackage{amssymb}
\usepackage{amsmath}
\usepackage{mathrsfs}
\usepackage{hyperref}
\usepackage{multirow}
\usepackage{dsfont}

\newcommand{\Beta}{\text{B}}

\title{$\alpha'$-Expansion of Open String Disk Integrals\\via Mellin Transformations}
\author{Ellis Ye Yuan}
\affiliation{Perimeter Institute for Theoretical Physics, Waterloo, ON N2L 2Y5,
Canada}
\affiliation{Department of Physics \& Astronomy, University of Waterloo, Waterloo, ON N2L 3G1,
Canada}
\emailAdd{yyuan@perimeterinstitute.ca}

\abstract{Open string disk integrals are represented as contour integrals of a product of Beta functions by using Mellin transformations. This makes the mathematical problem of computing the $\alpha'$ expansion around the field theory limit basically identical to that of the $\epsilon$ expansion in Feymann loop integrals around the four dimensional limit. More explicitly, the formula in Mellin space obtained directly from the standard Koba-Nielsen like representation is valid in a region of values of $\alpha'$ that does not include $\alpha'=0$. Analytic continuation is therefore needed since contours are pinched by poles as $\alpha'\to 0$. Deforming contours that get pinched by poles generates a set of $(n-3)!$ multidimensional residues left behind which contain all the field theory information. We end by drawing some analogies between the field theory formulas obtained by this method and those derived recently from using the scattering equations.}

\begin{document}
\maketitle

\section{Introduction}

The birth of string theory was witnessed by the magic Veneziano amplitude~\citep{Veneziano:1968yb}
\begin{equation}\label{venezianoamplitude}
A_4=\frac{\Gamma(\alpha's+\alpha_0)\,\Gamma(\alpha't+\alpha_0)}{\Gamma(\alpha'(s+t)+2\alpha_0)}=\Beta(\alpha's+\alpha_0,\alpha't+\alpha_0),
\end{equation}
which manifests a duality between the $s$ channel and the $t$ channel thanks to the presence of a Beta function. Its zero slope limit can be obtained directly by Laurent expansion of the Beta function. Especially, in the case $\alpha_0=0$, to the leading order this is as simple as
\begin{equation}\label{laurentexpansion}
\Beta(\alpha's,\alpha't)=\alpha'^{-1}\left(\frac{1}{s}+\frac{1}{t}\right)+\mathcal{O}(\alpha'^{0}).
\end{equation}
This formula has an integral representation
\begin{equation}\label{venezianointegral}
A_4=\int_{0}^{1}dz\,z^{\alpha's+\alpha_0}(1-z)^{\alpha't+\alpha_0},
\end{equation}
which has a higher multiplicity extension applied in the usual practice of calculating string amplitudes. In recent progress in exploring the structure of scattering among massless states in the open superstring in the pure spinor formalism, it was realized that the associated disk amplitude can be re-constructed by a field theory Kawai-Lewellen-Tye (KLT) relations combining ten-dimensional super Yang-Mills (SYM) amplitudes with the ``disk integrals'' $Z[\alpha|\beta]$~\citep{Mafra:2011nv,Mafra:2011nw,Broedel:2013tta}
\begin{equation}\label{openstringamplitude}
A^{\text{string}}(1,\ldots,n)=\sum_{\alpha,\beta\in S_{n-3}}Z[1,\ldots,n|1,\alpha,n,n-1]\,\mathcal{S}[\alpha|\beta]\,A^{\text{SYM}}(1,\beta,n-1,n),
\end{equation}
where $Z[\alpha|\beta]$ is defined as\footnote{Here we use $z_{a,b}$ instead of the usual $|z_{a,b}|$ in the Koba-Nielsen factor, since with a fixed ordering of $z$ this merely leads to a possible an overall sign, which is irrelevant in our discussion.}
\begin{equation}\label{diskdef}
Z[\alpha|\beta]:=\frac{1}{\text{vol }SL(2,\mathds{R})} \int_{z_{\alpha(1)}<\cdots<z_{\alpha(n)}} d^nz_a\, \frac{\prod_{a<b}z_{a,b}^{\alpha' s_{a,b}}}{z_{\beta(1),\beta(2)}\cdots z_{\beta(n-1),\beta(n)} z_{\beta(n),\beta(1)}},
\end{equation}
with $z_{a,b}=z_a-z_b$, and $s_{a,b}=2k_a\cdot k_b$ are the usual Mandelstam variables (since we consider only massless external states, $k^2_a=0$ for any $a$). $\mathcal{S}[\alpha|\beta]$ is the $(n-3)!\times(n-3)!$ based momentum kernel in its zero slope limit, which depends solely on the external data. The explicit definition of the momentum kernel can be found in \citep{BjerrumBohr:2010hn}, but this is irrelavent in this work. Based on the decomposition \eqref{openstringamplitude}, here we only focus on the disk integrals $Z$. These objects contain a universal Koba-Nielsen factor $\prod_{a<b}|z_a-z_b|^{\alpha's_{a,b}}$~\citep{Koba:1969rw}. In general the Koba-Nielsen factor has singular behavior when approaching the boundaries of the integration domain (or moduli space) of \eqref{diskdef}, which is responsible for the pole structure of the string amplitudes, but hinders a straightforward analytic evaluation. For example, already for \eqref{venezianoamplitude} when $\alpha_0\leq0$, outside the kinematic region where the integration \eqref{venezianointegral} is defined, one has to calculate at the cost of introducing a special contour in the complexified $z$ plane~\citep{2006JPhA...39.2509H,Witten:2013pra}. This problem still prevails in analyzing the leading terms in the zero slope limit, which captures the degeneration of the disk with marked points on the boundary into different tree-level Feyman diagrams of the corresponding field theory. Especially, for generic multiplicity the Koba-Nielsen factor possesses factors of the form $(z_a-z_b)$ that entangle different moduli. When extracting terms associated to a certain degenerated tree diagram, one has to take very good care of the relative separations between different pairs of adjacent points by introducing new parametrizations, in order to keep track of how the disk degenerates (see, e.g., \citep{Scherk:1971xy,Frizzo:1999zx}, and for a more recent closed string analog~\citep{Tourkine:2013rda}). By a careful study of the degeneration of the disk, in \cite{Mafra:2011nw} the authors proposed a set of extended variables, so that all different degeneration channels can be treated together.

Given these facts, it is worth to ask whether some method can automatically manipulates the singularities arising from the boundaries of the integration domain, but without caring much about how the boundaries look like. Such analysis may potentially give rise to a representation with some new geometrical understanding of string amplitudes, whose field-theory limit is organized in a way far different from the traditional Feynamn diagrams. As a first effort, let us make an observation that, the way of entangling different moduli in \eqref{diskdef} shares a very similar appearance with the form of any loop amplitude computation in ordinary field theory after introducing Feynman parameters. In the dimensional regularization of field theory, we have the well-developed Mellin-Barnes technique, whose core ingredient is the transformation~\citep{Smirnov:2004ym}
\begin{equation}\label{genericMellin}
(A_1+A_2)^\lambda=\int_{-i\infty}^{+i\infty}\frac{dw}{2\pi i}\,\Beta(-\lambda+w,-w)\,A_1^{\lambda-w}A_2^w.
\end{equation}
In this paper, we are going to show that the application of the Mellin transformation \eqref{genericMellin} in the disk integrals gives rise to a new integral representation involving Beta functions only, valid for all non-singular kinematic configurations. The original moduli are integrated out and we are left with integrations of the type arising from \eqref{genericMellin}, and the pole structure of the amplitude is now transmitted from the boundaries of the moduli space to the pole structure in the Mellin space, which can then be easily extracted by contour deformations. In particular, the leading terms (i.e., order $\alpha'^{3-n}$ for an $n$-particle scattering) are completely absorbed into the residues of the first poles picked up by this deformation, of which there are $(n-3)!$ in total (which is far less than the number of Feynman diagrams), and they are organized by a nice combinatorial structure. Since generically each of the $(n-3)!$ residues has spurious poles, this provides yet another non-local decomposition of the field theory amplitude, which is different from that of Britto-Cachazo-Feng-Witten (BCFW)~\citep{Britto:2004ap,Britto:2005fq}. As a glimpse of our results, the $\alpha'\to0$ limit of $Z[123456|123456]$ at $6$ points are completely absorbed into the following $6$ residues (Here we denote $s_{a:a+2}=s_{a,a+1}+s_{a,a+2}+s_{a+1,a+2}$)
\begin{equation}\label{6ptcanonical}
\begin{split}
\begin{split}
&\Beta(\alpha's_{2,3},\alpha'(s_{1,2}-s_{1:3}))\,\Beta(\alpha's_{2:4},\alpha'(s_{1:3}-s_{5,6}))\,\Beta(\alpha's_{1,6},\alpha's_{5,6})\\
+&\Beta(\alpha'(s_{3:5}-s_{4,5}),-\alpha'(s_{1,2}-s_{1:3}))\,\Beta(\alpha's_{2:4},\alpha'(s_{1,2}-s_{5,6}))\,\Beta(\alpha's_{1,6},\alpha's_{5,6})\\
+&\Beta(\alpha's_{3,4},-\alpha'(s_{3:5}-s_{4,5}))\,\Beta(\alpha's_{2:4},\alpha'(s_{1,2}-s_{5,6}))\,\Beta(\alpha's_{1,6},\alpha's_{5,6})\\
+&\Beta(\alpha's_{2,3},\alpha'(s_{1,2}-s_{1:3}))\,\Beta(\alpha's_{4,5},-\alpha'(s_{1:3}-s_{5,6}))\,\Beta(\alpha's_{1,6},\alpha's_{1:3})\\
+&\Beta(\alpha'(s_{3:5}-s_{4,5}),-\alpha'(s_{1,2}-s_{1:3}))\,\Beta(\alpha's_{4,5},-\alpha'(s_{1,2}-s_{5,6}))\,\Beta(\alpha's_{1,6},\alpha's_{1,2})\\
+&\Beta(\alpha's_{3,4},-\alpha'(s_{3:5}-s_{4,5}))\,\Beta(\alpha's_{3:5},-\alpha'(s_{1,2}-s_{5,6}))\,\Beta(\alpha's_{1,6},\alpha's_{1,2})
\end{split}
\end{split}
\end{equation}
as the leading terms at order $\alpha'^{-3}$ in its Laurent expansion \eqref{laurentexpansion} around $\alpha'=0$. There are altogether four spurious poles at $s_{1,2}-s_{1:3}=0$, $s_{3:5}-s_{4,5}=0$, $s_{1:3}-s_{5,6}=0$ and $s_{1,2}-s_{5,6}=0$ respectively, but they cancel away in the summation, resulting in $14$ terms corresponding to the $14$ planar scalar diagrams with the given ordering $(123456)$, as expected.

It is interesting that Mellin transformation has been an essential tool in the work of Stieberger and Taylor~\citep{Stieberger:2013hza,Stieberger:2013nha}. There it was motivated from the analogous structure of superstring and supergravity amplitudes. After switching the disk boundary positions $z_a$ to $n(n-3)/2$ cross-ratios $u_{a,b}=(z_a-z_b)(z_{a-1}-z_{b+1})/(z_a-z_{b+1})(z_{a-1}-z_{b})$ in one-to-one correspondence to the independent kinematic invariants $s_{a,b}$, Mellin transformation draws an exact correspondence between the two sides. In this paper Mellin transformation is used as the tool to disentangle $z$ variables and to manifest the singularity structure of general disk integrals in a different space.

The paper is organized as follows. In Section \ref{section2} we illustrate this method by discussing in great detail its application to the simpliest non-trivial case at five points. Following that, we generalize it to any multiplicity and any orderings in Section \ref{section3}. In Section \ref{section4} we show how to apply multi-dimensional contour deformation to extract the leading terms from the result in \ref{section2} in the zero slope limit, and give the general structure of these leading field-theory pieces obtained from this calculation. In the end we point out several possible future explorations.


\section{Five-Point Amplitude: Illustrating the Technique}\label{section2}

In this section we perform our analysis at $5$ points, which is the first non-trivial case. We choose the canonial ordering $I=(12\ldots n)$ on the boundary of the disk, and fix the $SL(2,\mathds{R})$ redundancy by setting $\{z_1,z_2,z_5\}=\{\infty,0,1\}$. As the first example we study $Z[I|I]$, i.e., letting the Parke-Taylor factor to share the same ordering as that on the boundary of the disk. In this case the amplitude is
\begin{equation}
\int dz_3dz_4\, z_3^{\alpha's_{2,3}-1}z_4^{\alpha's_{2,4}}(1-z_3)^{\alpha's_{3,5}}(1-z_4)^{\alpha's_{4,5}-1}(z_4-z_3)^{\alpha's_{3,4}-1},
\end{equation}
where the integration is carried out over the region $0\leq z_3\leq z_4\leq1$. We do a change of variables as
\begin{equation}
z_3=y_3y_4,\qquad z_4=y_4,
\end{equation}
so that both $y_3$ and $y_4$ are integrated over $[0,1]$, and the amplitude becomes
\begin{equation}
\int_0^1 dy_3dy_4\, y_3^{\alpha's_{2,3}-1}y_4^{\alpha's_{1,5}-1}(1-y_3y_4)^{\alpha's_{3,5}}(1-y_4)^{\alpha's_{4,5}-1}(1-y_3)^{\alpha's_{3,4}-1}.
\end{equation}
Our purpose is to fully integrate out the moduli $y_3$ and $y_4$, and so we need to decompose the factor $(1-y_3y_4)$ so that different moduli get separated. This is done using the Mellin transformation \eqref{genericMellin}. In order to avoid the appearance of any undesirable factor of the form $(-1)^w$ when applying \eqref{genericMellin}, we first choose to re-write this factor as
\begin{equation}
(1-y_3y_4)=(1-y_3)+y_3(1-y_4),
\end{equation}
i.e., setting $A_1=1-y_3$ and $A_2=y_3(1-y_4)$ in \eqref{genericMellin}. Then Mellin transformation gives
\begin{equation}\label{mellintransformation5pt}
(1-y_3y_4)^{\alpha's_{3,5}}=\int \frac{dw}{2\pi i}\, \Beta(-\alpha's_{3,5}+w,-w) (1-y_3)^{\alpha's_{3,5}-w}(y_3(1-y_4))^{w}.
\end{equation}
Plugging this into the above integrand, we obtain
\begin{equation}\label{5ptbetaintermediate}
\begin{split}
\int_{-i\infty}^{+i\infty} \frac{dw}{2\pi i}\, \Beta(-\alpha's_{3,5}+w,-w)&\cdot (\int_0^1dy_3\, y_3^{\alpha's_{2,3}-1+w}(1-y_3)^{\alpha'(s_{3,4}+s_{3,5})-1-w})\\
&\cdot (\int_0^1dy_4\, y_4^{\alpha's_{1,5}-1}(1-y_4)^{\alpha's_{4,5}-1+w}).
\end{split}
\end{equation}
At this point, we can directly complete the moduli integrations, and hence re-formulate this disk integral into the integration of a product of three Beta functions along a contour in the Mellin space $w$
\begin{equation}\label{5ptbetaamplitude}
\int_{-i\infty}^{+i\infty} \frac{dw}{2\pi i}\, \Beta(-\alpha's_{3,5}+w,-w)\Beta(\alpha's_{2,3}+w,\alpha'(s_{3,4}+s_{3,5})-w)\Beta(\alpha's_{1,5},\alpha's_{4,5}+w).
\end{equation}

The integrand in \eqref{5ptbetaamplitude} possesses $5$ families of poles, corresponding to
\begin{equation}\label{5ptpolefamilies}
w=\alpha's_{3,5}-m,\quad w=m,\quad w=-\alpha's_{2,3}-m,\quad w=\alpha'(s_{3,4}+s_{3,5})+m,\quad w=-\alpha's_{4,5}-m,
\end{equation}
where $m\in\mathds{N}$. Following the terminology in the literature we call the poles of the form $w=w^*-m$ left poles, and the poles of the form $w=w^*+m$ right poles~\citep{Smirnov:2004ym}. The contour in the integration \eqref{5ptbetaamplitude} should be chosen to sit between all left poles and all right poles and go from $-i\infty$ to $+i\infty$. To see this, notice that this is exactly the choice made in the Mellin transformation \eqref{mellintransformation5pt}, and is also the result of completing the $y$ integrations. So in determining the contour all Beta functions in the integrand should be taken into consideration.  In the simplest situation when the kinematic configuration satisfies
\begin{equation}
\max(s_{3,5},-s_{2,3},-s_{4,5})<\min(0,s_{3,4}+s_{3,5}),
\end{equation}
for any $\alpha'$ we are allowed to fix the contour to be merely a line parallel to the imaginary axis whose real part sits between the above two bounds, with which numerical integration can be trivially performed. In going from \eqref{5ptbetaintermediate} to \eqref{5ptbetaamplitude}, we directly translate the formal integral representation to the Beta functions themselves, and so we have avoided the problems arising from the kinematic region where the representation \eqref{5ptbetaintermediate} is not directly defined. Hence the new representation \eqref{5ptbetaamplitude} in the Mellin space is well-defined for generic kinematic configuration $\{k^\mu_a\}$ as long as it is non-singular.

With the formula in terms of Beta functions, the singularity that appears in the zero slope limit can be observed as coming from the failure in finding a contour satisfying the above criteria. In the explicit example \eqref{5ptbetaamplitude}, we see that as $\alpha'\rightarrow0$, poles corresponding to $m=0$ in all the families in \eqref{5ptpolefamilies} will collide at $w=0$, and so the original contour is not well-defined in the limit. However, since this singular behavior is only an effect of the first poles in each family, we can always extract this singular piece as residues at the colliding poles via contour deformation. We choose the convention to deform the contour rightwards. At $5$ points, from \eqref{5ptbetaamplitude} it is easy to see that when $\alpha'$ is so small that $|\alpha's_{a,b}|\ll1$ for any $s_{a,b}$, the contour can be deformed to a line parallel to the imaginary axis with its real part somewhere in $(0,1)$~\footnote{When $\alpha'$ sits in this region, the existence of this new contour is independent of kinematics $\{s_{a,b}\}$.}. During this deformation, we pick up two poles
\begin{equation}
w^*_1=0,\qquad w^*_2=\alpha'(s_{3,4}+s_{3,5}),
\end{equation}
whose residues are
\begin{equation}
\begin{split}
\text{Res}_{w=w^*_1}&=-\Beta(\alpha's_{2,3},\alpha'(s_{3,4}+s_{3,5}))\Beta(\alpha's_{1,5},\alpha's_{4,5}),\\
\text{Res}_{w=w^*_2}&=-\Beta(\alpha's_{3,4},-\alpha'(s_{3,4}+s_{3,5}))\Beta(\alpha's_{1,5},\alpha's_{1,2}).
\end{split}
\end{equation}
Hence for sufficiently small $\alpha'$ the disk integral becomes
\begin{equation}
\begin{split}
\Beta(\alpha's_{2,3},\alpha'(s_{3,4}+s_{3,5}))\Beta(\alpha's_{1,5},\alpha's_{4,5})+\Beta(\alpha's_{3,4},-\alpha'(s_{3,4}+s_{3,5}))\Beta(\alpha's_{1,5},\alpha's_{1,2})\\
+\int_{\frac{1}{2}-i\infty}^{\frac{1}{2}+i\infty}\frac{dw}{2\pi i}\, \Beta(-\alpha's_{3,5}+w,-w)\Beta(\alpha's_{2,3}+w,\alpha'(s_{3,4}+s_{3,5})-w)\Beta(\alpha's_{1,5},\alpha's_{4,5}+w),
\end{split}
\end{equation}
where the integration contour for $w$ is a line parallel to the imaginary axis with its real part chosen to be, say, $1/2$. Since each term in the above expression is well-defined in the limit, one can directly expand the formula around $\alpha'=0$ to obtain any order of $\alpha'$ correction. In particular, it can be easily seen that the leading order of the remaining integral is only $\alpha'^{-1}$, while the two residues are of order $\alpha'^{-2}$, and so the information of the zero slope limit is completely contained in the residues. In this example, the leading terms are
\begin{equation}
\begin{split}
-\text{Res}_{w=w^*_1}&\longrightarrow\alpha'^{-2}\left(\frac{1}{s_{2,3}}+\frac{1}{s_{3,4}+s_{3,5}}\right)\left(\frac{1}{s_{1,5}}+\frac{1}{s_{4,5}}\right),\\
-\text{Res}_{w=w^*_2}&\longrightarrow\alpha'^{-2}\left(\frac{1}{s_{3,4}}-\frac{1}{s_{3,4}+s_{3,5}}\right)\left(\frac{1}{s_{1,5}}+\frac{1}{s_{1,2}}\right).
\end{split}
\end{equation}
At first sight, there is an unphysical pole at $s_{3,4}+s_{3,5}=0$, but as we sum up the two residues this spurious pole cancels out and we are left with the correct planar amplitude in the colored $\phi^3$ theory (apart from the factor $\alpha'^{-2}$)
\begin{equation}
\frac{1}{s_{1,2}s_{3,4}}+\frac{1}{s_{2,3}s_{4,5}}+\frac{1}{s_{3,4}s_{1,5}}+\frac{1}{s_{4,5}s_{1,2}}+\frac{1}{s_{1,5}s_{2,3}}.
\end{equation}
Hence we see that the Mellin transformation provides a decomposition of the $5$-pt planar scalar amplitude into $2$ terms, each of which is non-local. Roughly speaking, this non-locality originates from distributing the information associated to the $n(n-3)/2$ channels consistent with the disk into $(n-3)$ moduli via Mellin transformation, with the purpose of regarding the moduli space after the prescribed gauge-fixing as a direct product of $(n-3)$ $1$-dimensional space.


The same analysis can be done for any other orderings of the Parke-Taylor factor. We postpone the details of the derivation to the next section in the discussion with the general setting. Here we merely give some convenient notations and show the result. First we denote $s_{a:a+m}=(k_a+k_{a+1}+\cdots+k_{a+m})^2$. Since we only consider massless scattering, we always have $s_{a:b}=\sum_{a\leq c<d\leq b}s_{c,d}$. We also define the symbol $\theta_{a,b}=1$ if the labels $a$ and $b$ are adjacent in the Parke-Taylor factor and $0$ otherwise, and in the same spirit we abbreviate $\theta_{a:b}=\sum_{a\leq c<d\leq b}\theta_{c,d}$. It is very convenient to use the notation
\begin{equation}\label{shat}
\hat{s}_{a,b}=\alpha's_{a,b}-\theta_{a,b}.
\end{equation}
With this definition, the $5$-pt formula for any ordering takes the form
\begin{equation}\label{5ptamplitudegeneral}
\int dz_3dz_4\, z_3^{\hat{s}_{2,3}}z_4^{\hat{s}_{2,4}}(1-z_3)^{\hat{s}_{3,5}}(1-z_4)^{\hat{s}_{4,5}}(z_4-z_3)^{\hat{s}_{3,4}},
\end{equation}
where due to the Parke-Taylor factor origin of $\{\theta_{a,b}\}$ we also require (for any $a$)
\begin{equation}
\sum_{b\neq a}\theta_{a,b}=2.
\end{equation}

Using the same Mellin transformation, \eqref{5ptamplitudegeneral} can be transformed to
\begin{equation}
\int \frac{dw}{2\pi i}\, \Beta(-\hat{s}_{3,5}+w,-w)\Beta(1+\hat{s}_{2,3}+w,1+\hat{s}_{3,4}+\hat{s}_{3,5}-w)\Beta(2+\hat{s}_{2:4},1+\hat{s}_{4,5}+w).
\end{equation}
The pattern of colliding poles depends on specific orderings in general. To be explicit, in the $\alpha'\rightarrow0$ limit the first pole in each of the five families of poles now become
\begin{equation}
\begin{split}
\text{left poles:}&\quad w=-\theta_{3,5},\quad w=\theta_{2,3}-1,\quad w=\theta_{4,5}-1,\\
\text{right poles:}&\quad w=0,\quad w=-\theta_{3,4}-\theta_{3,5}+1.
\end{split}
\end{equation}
From the definition of $\theta$, we know that $-\theta_{3,4}-\theta_{3,5}+1$ can only choose value among $\{-1,0,1\}$. When it is $-1$, we are forced to have $\theta_{3,4}=\theta_{3,5}=1$ and $\theta_{2,3}=\theta_{4,5}=0$, and so
\begin{equation}\label{poleboundsid}
\max(-\theta_{3,5},\theta_{2,3}-1,\theta_{4,5}-1)=\min(0,-\theta_{3,4}-\theta_{3,5}+1).
\end{equation}
If on the other hand $-\theta_{3,4}-\theta_{3,5}+1\geq0$, since it is obvious that the first pole in any of the three left families can never be greater than $0$, the identity \eqref{poleboundsid} is still valid. Hence we see that at $5$ points, whatever the ordering of Parke-Taylor factor is, the singular behavior is only a result of collision among the first poles from the left and the right families, and so in order to fully resolve it we only need to deform the contour rightwards to extract one of or both of the two first right poles (depending on whether the two right families become identical in the limit, i.e., whether $-\theta_{3,4}-\theta_{3,5}+1=0$), which are
\begin{equation}
w^*_1=0,\quad w^*_2=1+\hat{s}_{3,4}+\hat{s}_{3,5},
\end{equation}
whose residues are
\begin{equation}
\begin{split}
\text{Res}_{w=w^*_1}&=-\Beta(1+\hat{s}_{2,3},1+\hat{s}_{3,4}+\hat{s}_{3,5})\Beta(2+\hat{s}_{2:4},1+\hat{s}_{4,5}),\\
\text{Res}_{w=w^*_2}&=-\Beta(1+\hat{s}_{3,4},-1-\hat{s}_{3,4}-\hat{s}_{3,5})\Beta(2+\hat{s}_{2:4},2+\hat{s}_{3:5}).
\end{split}
\end{equation}
Actually, if we are only interested in the leading terms in the $\alpha'\rightarrow0$ limit, this can always be directly extracted from the $\alpha'$ expansion of the summation of the two residues
\begin{equation}\label{5ptleadingterms}
\Beta(1+\hat{s}_{2,3},1+\hat{s}_{3,4}+\hat{s}_{3,5})\Beta(2+\hat{s}_{2:4},1+\hat{s}_{4,5})+\Beta(1+\hat{s}_{3,4},-1-\hat{s}_{3,4}-\hat{s}_{3,5})\Beta(2+\hat{s}_{2:4},2+\hat{s}_{3:5}),
\end{equation}
regardless of specific ordering of the Parke-Taylor factor (since when one pole does not contribute, it just means that its residue vanishes at the leading order). So in general the leading behavior consists of two pieces as in \eqref{5ptleadingterms}. As in the case of canonical ordering, each piece may contain a spurious pole $1/(s_{3,4}+s_{3,5})$, but it always cancels and the summation reduces exactly to the corresponding field theory diagrams.

In fact, due to the simplicity of just a single contour integration at $5$ points, for any value of $\alpha'$ we can straightforwardly push the contour all the way to the real infinity, picking up all the poles in the two right families, which transforms the $5$-pt formula into a series expansion
\begin{equation}\label{infiniteseries}
\begin{split}
\sum_{m=0}^{\infty}\big(
&\Beta(m+1+\hat{s}_{2,3},-m+1+\hat{s}_{3,4}+\hat{s}_{3,5})\Beta(2+\hat{s}_{2:4},m+1+\hat{s}_{4,5})\\
&+\Beta(m+1+\hat{s}_{3,4},-m-1-\hat{s}_{3,4}-\hat{s}_{3,5})\Beta(2+\hat{s}_{2:4},m+2+\hat{s}_{3:5})
\big).
\end{split}
\end{equation}

The method discussed in this section can be conveniently extended to higher multiplicities and any orderings, as we are going to see in the next section. Although for higher multiplicities we will inevitably encounter multidimensional contour integrations that complicate the contour deformation, the leading behavior of the disk integral in the $\alpha'\rightarrow0$ limit remains relatively simple, since it is still fully absorbed in the first poles picked up by the deformation. This provides a new way of decomposing the field theory counterpart.


\section{Generic Disk Integrals in Terms of Beta Functions}\label{section3}

In this section we generalize the above discussion to any multiplicities and any orderings (still with the canonical ordering on the disk boundary), following a particular way of doing Mellin transformations. The redundancy in the disk integration is always fixed by setting $\{z_1,z_2,z_n\}=\{-\infty,0,1\}$, and for any ordering in the Parke-Taylor factor the integrand is
\begin{equation}\label{originalintegrand}
\prod_{a=3}^{n-1}z_a^{\hat{s}_{2,a}}(1-z_a)^{\hat{s}_{a,n}}\prod_{3\leq a<b\leq n-1}(z_b-z_a)^{\hat{s}_{a,b}}.
\end{equation}
We generalize the transformation to $y$ variables by
\begin{equation}\label{changeofvariablegeneral}
z_a=\prod_{b=a}^{n-1}y_b,\quad 3\leq a\leq n-1,
\end{equation}
so to bring the integration domain for every $y_a$ to be $[0,1]$. With the notations introduced before as well as $y_{a:b}=y_ay_{a+1}\cdots y_b$, the new integrand can be conveniently expressed as
\begin{equation}
\prod_{a=3}^{n-1}y_a^{(a-3)+\hat{s}_{2:a}}(1-y_a)^{\hat{s}_{a,a+1}}\prod_{3\leq a<b\leq n-1}(1-y_{a:b})^{\hat{s}_{a,b+1}}.
\end{equation}

There are in total $(n-3)(n-4)/2$ factors in the form $(1-y_{a:b})$. We use Mellin transformation to resolve these factors into products of $y_a$'s and $(1-y_a)$'s. As before, in order to avoid the situation of producing factors like $(-1)^w$, this has to be done step by step. With the given change of variables \eqref{changeofvariablegeneral}, in each step it is most natural to re-arrange the factor as
\begin{equation}
(1-y_{a:b})=(1-y_a)+y_a(1-y_{a+1:b}),
\end{equation}
to separate $y_a$ from the rest. In this case the corresponding Mellin transformation is
\begin{equation}\label{mellintransformation}
(1-y_{a:b})^{p_{a,b}}=\int \frac{dw_{a,b}}{2\pi i} \Beta(-p_{a,b}+w_{a,b},-w_{a,b}) (1-y_a)^{p_{a,b}-w_{a,b}}(y_a(1-y_{a+1:b}))^{w_{a,b}}.
\end{equation}
Here the subscript $\{a,b\}$ in $w_{a,b}$ is just a label for the new variable to keep track of its origin. It is easy to see from the above that one has to transform $(1-y_{a-1:b})$ before $(1-y_{a:b})$ for any fixed $b$, and the power $p_{a,b}$ of the latter will receive a contribution from the transformation of the former, i.e.,
\begin{equation}
p_{a,b}=\hat{s}_{a,b+1}+(1-\delta_{3,a})w_{a-1,b},
\end{equation}
where $\delta_{a,b}$ is the usual Kronecker delta. After all the transformations are performed, the $y$-dependant part of the integrand becomes
\begin{equation}
\prod_{3\leq a\leq n-1}
\left(y_a^{(a-3)+\hat{s}_{2:a}+\sum_{b=a+1}^{n-1}w_{a,b}}(1-y_a)^{\hat{s}_{a,a+1}+\sum_{c=a+1}^{n-1}w_{a,c}(p_{a,c}-w_{a,c})+(1-\delta_{3,a})w_{a-1,a}}\right).
\end{equation}
After all the moduli are integrated out, we are left with contour integrations of purely a product of Beta functions over the $(n-3)(n-4)/2$ $w_{a,b}$ variables (using abbreviation $\hat{\delta}_{a,b}=1-\delta_{a,b}$)
\begin{equation}\label{betaamplitudesgeneral}
\begin{split}
\int&\prod_{3\leq a<b\leq n-1}\frac{dw_{a,b}}{2\pi i}
\prod_{3\leq a<b\leq n-1}\Beta_{a,b}(-\hat{s}_{a,b+1}-\hat{\delta}_{3,a}w_{a-1,b}+w_{a,b},-w_{a,b})\cdot\\
&\quad\cdot\prod_{a=3}^{n-1}\Beta_{a}\left((a-2)+\hat{s}_{2:a}+\sum_{c=a+1}^{n-1}w_{a,c},1+\sum_{c=a}^{n-1}(\hat{s}_{a,c+1}+\hat{\delta}_{3,a}w_{a-1,c})-\sum_{c=a+1}^{n-1}w_{a,c}\right).
\end{split}
\end{equation}
For the convenience of later discussions, here we also put additional subscripts to the Beta function to keep track of their origins. The Beta functions labeled by two subscripts come from the Mellin transformations and are in one-to-one correspondence with the $w$ variables, while those labeled by a single subscript arise from completing the integrations of the original moduli.

In particular, if we just focus on $Z[I|I]$ which leads to the summation of all color-ordered planar scalar diagrams, we have $\theta_{a,b+1}=0$ for any $a<b$, and $\theta_{2:a}=a-2$ and $\sum_{b=a}^{n-1}\hat{\theta}_{a,b+1}=\theta_{a,a+1}=1$. So the above integration reduces to
\begin{equation}\label{betaamplitudescanonical}
\begin{split}
\int&\prod_{3\leq a<b\leq n-1}\frac{dw_{a,b}}{2\pi i}
\prod_{3\leq a<b\leq n-1}\Beta_{a,b}(-\alpha's_{a,b+1}-\hat{\delta}_{3,a}w_{a-1,b}+w_{a,b},-w_{a,b})\cdot\\
&\quad\cdot\prod_{a=3}^{n-1}\Beta_{a}\left(\alpha's_{2:a}+\sum_{c=a+1}^{n-1}w_{a,c},\sum_{c=a}^{n-1}(\alpha's_{a,c+1}+\hat{\delta}_{3,a}w_{a-1,c})-\sum_{c=a+1}^{n-1}w_{a,c}\right).
\end{split}
\end{equation}

The contours of the integrations are constrained by all the Beta functions in the same way as the $5$-pt case. To be more explicit, when we focus on a certain variable $w_{a,b}$, the integrand in \eqref{betaamplitudesgeneral} will impose poles of two types in the $w_{a,b}$ plane
\begin{align}
\text{left poles:}\quad w_{a,b}&=-m+F(\alpha',\{s\},\{w\}\backslash\{w_{a,b}\}),\quad m\in\mathbb{N}\\
\text{right poles:}\quad w_{a,b}&=+m+G(\alpha',\{s\},\{w\}\backslash\{w_{a,b}\}),\quad m\in\mathbb{N},
\end{align}
where $F$ and $G$ are functions that can be read from the arguments in the Beta functions (there can be several different $F$'s and $G$'s, each of which produces a series of poles). The contour for $w_{a,b}$ integration is chosen from $-i\infty$ to $+i\infty$ such that it separates all the left poles from all the right poles.


\section{Contour Deformation and the Leading-Order Singularities}\label{section4}

\subsection{General Discussions}

As with the $5$-pt case, for higher multiplicities the singular behavior of the disk integral in the $\alpha'\rightarrow0$ limit also comes from the failure in finding a well-defined contour, due to the collision of the left and right poles. Generically we cannot expect that a contour deformation simply passing only all the first poles in the right families will make the remaining integrations completely well-defined, since the collision may involve poles other than the first ones. However, as long as the original contour becomes ill-defined, at least a subset of the first poles will get pinched with some other poles, and by passing these poles in contour deformation the singular behavior of the integration must at least be partially resolved, and so the terms of the leading order $\alpha'^{3-n}$ must all be absorbed into the residues of the first poles.

Here we determine where the first poles of the right families locate. First we need to fix a certain ordering in integrating out the $w$ variables. Here we make a special choice
\begin{equation}\label{integrationordering}
\int dw_{3,4}\cdots\int dw_{3,n-1}\int dw_{4,5}\cdots\int d_{n-3,n-2}\int d_{n-3,n-1}\int dw_{n-2,n-1}\,\mathcal{I}_{\text{tot}},
\end{equation}
where $\mathcal{I}_{\text{tot}}$ denotes the total integrand. The poles of the Beta functions directly shown in the integrand, when expressed as conditions on any $w_{a,b}$, are
\begin{align}
\label{poles1}w_{a,b}&=m,\\
\label{poles2}w_{a,b}&=-m+\hat{s}_{a,b+1}+\hat{\delta}_{3,a}w_{a-1,b},
\end{align}
and for $b=n-1$ we have two additional series
\begin{align}
\label{poles3}w_{a,n-1}&=m+1+\sum_{c=a}^{n-1}(\hat{s}_{a,c+1}+\hat{\delta}_{3,a}w_{a-1,c})-\sum_{c=a+1}^{n-2}w_{a,c},\\
\label{poles4}w_{a,n-1}&=-m-(a-2)-\hat{s}_{2:a}-\sum_{c=a+1}^{n-2}w_{a,c}.
\end{align}
Note here that when a constraint of above types contains several variables, it should be understood as constraining the variable that comes right-most in the ordering \eqref{integrationordering}.

From \eqref{poles1}, \eqref{poles2}, \eqref{poles3} and \eqref{poles4}, we can observe that the locations of possible poles for $w_{a,b}$ is independent of the value of any $w_{a',b'}$ with $a'>a$ and $b'$ arbitrary. So when we study the set of variables $\{w_{a^*,b}\}$ for any fixed $a^*$ (in total $n-a^*-1$ variables), we can ignore all the integrations sitting on the right. To determine the poles that we need to pick up for this set, only the Beta functions $B_{a^*}$ and $\{B_{a^*,b}\}$ are needed, altogether $n-a^*$ of them. Hence the entire integrand has the natural decomposition
\begin{equation}\label{integrandbstar}
\mathcal{I}_{\text{tot}}=\prod_{a^*=3}^{n-1}\mathcal{I}_{a^*},\quad\text{where}\quad\mathcal{I}_{a^*}=\Beta_{a^*}\prod_{b=a^*+1}^{n-1}\Beta_{a^*,b}.
\end{equation}
And in determining the locations of the first poles, we can pretend to study the poles of each $\mathcal{I}_{a^*}$ for a fixed $a^*$ ($a^*<n-1$), regarded as living in the $\{w_{a^*,b}\}$-space. These gives the $\{w_{a^*,b}\}$ coordinates of the poles of $\mathcal{I}_{\text{tot}}$. The remaining $\mathcal{I}_{n-1}$ is just a single Beta function $\Beta_{n-1}$, which never plays any role in determining the right poles. In the end, each pole of $\mathcal{I}_{\text{tot}}$ is obtained by all possible combinations of the $\{w_{a^*,b}\}$ values from each $a^*$.


Now we focus on a fixed $a^*$. Although all the right poles accompanying contour deformations seem to come from only \eqref{poles1} and \eqref{poles3}, we should be careful since they can also emerge from \eqref{poles2} for $w_{a^*,b}$ with $b<n-1$. The deformation of $w_{a^*,n-1}$ picks up two poles directly by \eqref{poles1} and \eqref{poles3}
\begin{align}
\label{polew3bA}w_{a^*,n-1}&=0,\\
\label{polew3bB}w_{a^*,n-1}&=1+\sum_{c=a}^{n-1}(\hat{s}_{a,c+1}+\hat{\delta}_{3,a}w_{a-1,c})-\sum_{c=a+1}^{n-2}w_{a,c}.
\end{align}

Firstly, when the isolated pole satisfies \eqref{polew3bA}, since the constraint that leads to \eqref{polew3bB} is regarded as determining the right pole of $w_{a^*,n-1}$ only, it can no longer be used to determining right poles for any variables that come later. So we are left with only one choice of right pole for $w_{a^*,n-2}$, corresponding to \eqref{poles1}. Secondly, when the isolated pole satisfies \eqref{polew3bB}, apart from the choice \eqref{poles1} for $w_{a^*,n-2}$, if we substitute the value for $w_{a^*,n-1}$ into \eqref{poles2}, an additional family of right poles for $w_{a^*,n-2}$ is emerged
\begin{equation}\label{polew4bB}
w_{a,n-2}=m+1+\sum_{c=a}^{n-2}(\hat{s}_{a,c+1}+\hat{\delta}_{3,a}w_{a-1,c})-\sum_{c=a+1}^{n-3}w_{a,c}.
\end{equation}
Since \eqref{poles2} was not used to determined the right pole of $w_{a^*,n-2}$, the emergent condition \eqref{polew4bB} does determine a pole, and so in this case we get two poles for $w_{a^*,n-2}$.

By applying the above arguments recursively, it is easy to verify that in general the actual poles that one will encounter in deforming the $w_{a^*,b}$ contour are
\begin{align}
\label{polegeneralA}w_{a^*,b}&=0,\\
\label{polegeneralB}w_{a^*,b}&=1+\sum_{c=a}^{b}(\hat{s}_{a,c+1}+\hat{\delta}_{3,a}w_{a-1,c})-\sum_{c=a+1}^{b-1}w_{a,c}.
\end{align}
When the pole of $w_{a^*,b^*}$ is specified by \eqref{polegeneralA} for any $b^*$, the pole of $w_{a^*,b^*-1}$ is uniquely determined by \eqref{polegeneralA}. When it is specified by \eqref{polegeneralB} instead, the pole of $w_{a^*,b^*-1}$ has both choices \eqref{polegeneralA} and \eqref{polegeneralB}. Hence a simple counting shows that the contour deformations produce $n-a^*$ isolated poles for the variable set $\{w_{a^*,b}\}$ of a fixed $a^*$ (when ignoring the remaining variables), and thus when taking all $a^*$ into consideration we obtain
\begin{equation}
\prod_{a^*=3}^{n-2}(n-a^*)=(n-3)!
\end{equation}
isolated poles in total.

To obtain the residues upon these isolated poles, as a result of the natural decomposition \eqref{integrandbstar}, similar to the discussion of pole locations we can start by studying the residue of a single $\mathcal{I}_{a^*}$ with fixed $a^*$ regarded as a function of the variables $\{w_{a^*,b}\}$ only. The residue thus obtained will in general involve remaining variables $\{w_{a^*-1,b}\}$, which is left abitrary at this stage since they play no roles in $\mathcal{I}_{a^*}$. Later on as we assemble this residue with that from $\mathcal{I}_{a^*-1}$, these remaining variables should be understood as taking the corresponding values in determining the pole of $\mathcal{I}_{a^*-1}$ (this will be illustrated by an explicit example at $6$ points in the next subsection).

For a fixed $a^*$, if the pole is specified by \eqref{polegeneralB} for variables $w_{a^*,b}$ from $b=n-1$ all the way to $b=b^*$, then the Beta functions that has been used in picking up these poles are $\{\Beta_{a^*},\Beta_{a^*,n-1},\ldots,\Beta_{a^*,b^*+1}\}$. They altogether produces a factor $(-1)^{n-b^*}$, and after performing these integrations this part of the integrand reduces to
\begin{equation}
\begin{split}
(-1)^{n-b^*}&\prod_{b=a^*+1}^{b^*-1}\Beta_{a^*,b}(-\hat{s}_{a^*,b+1}-\hat{\delta}_{3,a^*}w_{a^*-1,b}+w_{a^*,b},-w_{a^*,b})\cdot\\
\cdot&\Beta_{a^*,b^*}(1+\sum_{c=a^*}^{b^*-1}(\hat{s}_{a^*,c+1}+\hat{\delta}_{3,a^*}w_{a^*-1,c})-\sum_{c=a^*+1}^{b^*-1}w_{a^*,c},\\
&\quad\quad\quad-1-\sum_{c=a^*}^{b^*}(\hat{s}_{a^*,c+1}+\hat{\delta}_{3,a^*}w_{a^*-1,c})+\sum_{c=a^*+1}^{b^*-1}w_{a^*,c})
\end{split}.
\end{equation}
When the remaining $w_{a^*,b}$ are specified by \eqref{polegeneralA}, since they come from $B_{a^*,b}$ for $b<b^*$, in the end we get the residue
\begin{equation}\label{residueA}
(-1)^{n-a^*-1}\Beta_{a^*,b^*}(1+\sum_{c=a^*}^{b^*-1}(\hat{s}_{a^*,c+1}+\hat{\delta}_{3,a^*}w_{a^*-1,c}),-1-\sum_{c=a^*}^{b^*}(\hat{s}_{a^*,c+1}+\hat{\delta}_{3,a^*}w_{a^*-1,c}))
\end{equation}
where the remaining $\{w_{a^*-1,b}\}$ are understood to be substituted by the pole locations determined from later integrations. Moreover, in the extreme case where all $\{w_{a^*,b}\}$ are set to zero, we have instead
\begin{equation}\label{residueB}
(-1)^{n-a^*-1}\Beta_{a^*}((a^*-2)+\hat{s}_{2:a^*},1+\sum_{c=a^*}^{n-1}(\hat{s}_{a^*,c+1}+\hat{\delta}_{3,a^*}w_{a^*-1,c})).
\end{equation}

To this end, we are able to assemble the results in \eqref{residueA} and \eqref{residueB} for different $a^*$'s together to get the full expression of residues at each of the $(n-3)!$ first poles of $\mathcal{I}_{\text{tot}}$. Their summation is
\begin{equation}
\alpha'^{3-n}\sum_{\{w^*_{a,b}\}}\left[(\mathcal{I}_{n-1}|_{w_{n-2,n-1}=w^*_{n-2,n-1}})\prod_{a^*=3}^{n-2}(\text{Res}(\mathcal{I}_{a^*};w_{a^*,b}=w^*_{a^*,b})|_{w_{a^*-1,b}=w^*_{a^*-1,b}})\right],
\end{equation}
where $\text{Res}(\mathcal{I};\{w\}=\{w^*\})$ denotes the residue of $\mathcal{I}$ at the pole specified by $\{w\}=\{w^*\}$ (i.e., \eqref{residueA} or \eqref{residueB}), and the summation is over all $(n-3)!$ poles, which are specified by the values $\{w_{a,b}\}=\{w^*_{a,b}\}$.


\subsection{Six-Point Amplitudes}

To illustrate the above discussion, here we provide the explicit result for a generic $6$-pt amplitude. The integrand has the decomposition $\mathcal{I}=\mathcal{I}_3\mathcal{I}_4\mathcal{I}_{5}$, where
\begin{equation}
\begin{split}
\mathcal{I}_{3}=&\Beta_{3}(1+\hat{s}_{2,3}+w_{3,4}+w_{3,5},1+\hat{s}_{3,4}+\hat{s}_{3,5}+\hat{s}_{3,6}-w_{3,4}-w_{3,5})\\
&\cdot\Beta_{3,4}(-\hat{s}_{3,5}+w_{3,4},-w_{3,4})\Beta_{3,5}(-\hat{s}_{3,6}+w_{3,5},-w_{3,5}),\\
\mathcal{I}_{4}=&\Beta_4(2+\hat{s}_{2:4}+w_{4,5},1+\hat{s}_{4,5}+\hat{s}_{4,6}+w_{3,4}+w_{3,5}-w_{4,5})\\
&\cdot\Beta_{4,5}(-\hat{s}_{4,6}-w_{3,5}+w_{4,5},-w_{4,5}),\\
\mathcal{I}_{5}=&\Beta_5(3+\hat{s}_{2:5},1+\hat{s}_{5,6}+w_{4,5}).
\end{split}
\end{equation}
As was shown in the previous discussion, the $\{w_{a^*,b}\}$ coordinates (with a given $a^*$) for any pole are determined only by $\mathcal{I}_{a^*}$. On the one hand, $w_{4,5}$ picks up two poles, upon which we have
\begin{equation}\label{6ptA}
\begin{split}
&\text{Res}(\mathcal{I}_4;w_{4,5}=0)=-\Beta_4(2+\hat{s}_{2:4},1+\hat{s}_{4,5}+\hat{s}_{4,6}+w_{3,4}+w_{3,5}),\\
&\text{Res}(\mathcal{I}_4;w_{4,5}=1+\hat{s}_{4,5}+\hat{s}_{4,6}+w_{3,4}+w_{3,5})\\
&\qquad\qquad\qquad\quad=-\Beta_{4,5}(1+\hat{s}_{4,5}+w_{3,4},-1-\hat{s}_{4,5}-\hat{s}_{4,6}-w_{3,4}-w_{3,5}).
\end{split}
\end{equation}
On the other hand, $\{w_{3,4},w_{3,5}\}$ together picks up $3$ poles, whose residues are
\begin{equation}\label{6ptB}
\begin{split}
&\text{Res}(\mathcal{I}_3;w_{3,4}=0,w_{3,5}=0)=\Beta_3(1+\hat{s}_{2,3},1+\hat{s}_{3,4}+\hat{s}_{3,5}+\hat{s}_{3,6}),\\
&\text{Res}(\mathcal{I}_3;w_{3,4}=0,w_{3,5}=1+\hat{s}_{3,4}+\hat{s}_{3,5}+\hat{s}_{3,6}-w_{3,4})\\
&\qquad\qquad\qquad\qquad\qquad\quad=\Beta_{3,5}(1+\hat{s}_{3,4}+\hat{s}_{3,5},-1-\hat{s}_{3,4}-\hat{s}_{3,5}-\hat{s}_{3,6}),\\
&\text{Res}(\mathcal{I}_3;w_{3,4}=1+\hat{s}_{3,4}+\hat{s}_{3,5},w_{3,5}=1+\hat{s}_{3,4}+\hat{s}_{3,5}+\hat{s}_{3,6}-w_{3,4})\\
&\qquad\qquad\qquad\qquad\qquad\quad=\Beta_{3,4}(1+\hat{s}_{3,4},-1-\hat{s}_{3,4}-\hat{s}_{3,5}).
\end{split}
\end{equation}
So the six poles are specified by
\begin{equation}
\begin{array}{cccc}
\begin{split}
\text{pole 1:}&\quad\\
\text{pole 2:}&\quad\\
\text{pole 3:}&\quad\\
\text{pole 4:}&\quad\\
\text{pole 5:}&\quad\\
\text{pole 6:}&\quad\\
\end{split}
&
\begin{split}
w_{3,4}&=0,\\
w_{3,4}&=0,\\
w_{3,4}&=1+\hat{s}_{3,4}+\hat{s}_{3,5},\\
w_{3,4}&=0,\\
w_{3,4}&=0,\\
w_{3,4}&=1+\hat{s}_{3,4}+\hat{s}_{3,5},
\end{split}
&
\begin{split}
w_{3,5}&=0,\\
w_{3,5}&=1+\hat{s}_{3,4}+\hat{s}_{3,5}+\hat{s}_{3,6},\\
w_{3,5}&=\hat{s}_{3,6},\\
w_{3,5}&=0,\\
w_{3,5}&=1+\hat{s}_{3,4}+\hat{s}_{3,5}+\hat{s}_{3,6},\\
w_{3,5}&=\hat{s}_{3,6},
\end{split}
&
\begin{split}
w_{4,5}&=0;\\
w_{4,5}&=0;\\
w_{4,5}&=0;\\
w_{4,5}&=1+\hat{s}_{4,5}+\hat{s}_{4,6};\\
w_{4,5}&=2+\hat{s}_{3:6}-\hat{s}_{5,6};\\
w_{4,5}&=2+\hat{s}_{3:6}-\hat{s}_{5,6}.
\end{split}
\end{array}
\end{equation}
The entire result is obtained by taking the product of $B_5$ with all possible choices of residues in \eqref{6ptA} and \eqref{6ptB}, with the variable $w_{4,5}$ in $B_5$ substituted by the value corresponding to the choice from \eqref{6ptA} it multiplies, and the variables $\{w_{3,4},w_{3,5}\}$ in \eqref{6ptA} substituted by the values corresponding to the choice from \eqref{6ptB} it multiplies. Hence, the total contribution from the first poles is a summation of six terms
%
%
%
\begin{equation}\label{6ptint0}
\begin{split}
&\Beta_3(1+\hat{s}_{2,3},1+\hat{s}_{3,4}+\hat{s}_{3,5}+\hat{s}_{3,6})\Beta_4(2+\hat{s}_{2:4},1+\hat{s}_{4,5}+\hat{s}_{4,6})\Beta_5(3+\hat{s}_{2:5},1+\hat{s}_{5,6})\\
+&\Beta_{3,5}(1+\hat{s}_{3,4}+\hat{s}_{3,5},-1-\hat{s}_{3,4}-\hat{s}_{3,5}-\hat{s}_{3,6})\Beta_4(2+\hat{s}_{2:4},2+\hat{s}_{3:6}-\hat{s}_{5,6})\Beta_5(3+\hat{s}_{2:5},1+\hat{s}_{5,6})\\
+&\Beta_{3,4}(1+\hat{s}_{3,4},-1-\hat{s}_{3,4}-\hat{s}_{3,5})\Beta_4(2+\hat{s}_{2:4},2+\hat{s}_{3:6}-\hat{s}_{5,6})\Beta_5(3+\hat{s}_{2:5},1+\hat{s}_{5,6})\\
+&\Beta_3(1+\hat{s}_{2,3},1+\hat{s}_{3,4}+\hat{s}_{3,5}+\hat{s}_{3,6})\Beta_{4,5}(1+\hat{s}_{4,5},-1-\hat{s}_{4,5}-\hat{s}_{4,6})\Beta_5(3+\hat{s}_{2:5},2+\hat{s}_{4:6})\\
+&\Beta_{3,5}(1+\hat{s}_{3,4}+\hat{s}_{3,5},-1-\hat{s}_{3,4}-\hat{s}_{3,5}-\hat{s}_{3,6})\Beta_{4,5}(1+\hat{s}_{4,5},-2-\hat{s}_{3:6}+\hat{s}_{5,6})\Beta_5(3+\hat{s}_{2:5},3+\hat{s}_{3:6})\\
+&\Beta_{3,4}(1+\hat{s}_{3,4},-1-\hat{s}_{3,4}-\hat{s}_{3,5})\Beta_{4,5}(2+\hat{s}_{3:5},-2-\hat{s}_{3:6}+\hat{s}_{5,6})\Beta_5(3+\hat{s}_{2:5},3+\hat{s}_{3:6}).
\end{split}
\end{equation}
For the canonical ordering, this reduces to \eqref{6ptcanonical} shown in the introduction. As in the $5$-pt case, the leading terms in the zero slope limit of this disk integral can again be obtain by direct Laurent expansion of these six residues.

\section{Discussions}

In this paper we have demonstrated that the leading singular terms in the zero slope limit of open string disk integrals can be directly extracted from Mellin transformations and contour deformations, in similar way as one usually does in evaluating field theory loop diagrams. As a result, the disk integrals can be fully expressed in terms of Beta functions. Generically these Beta functions give rise to several families of poles in the Mellin space, and terms at leading order $\alpha'^{3-n}$ are fully captured by the residues of the first poles in the right (or left) families. With our choice of procedure in doing Mellin transformations and deforming the integration contours, there are $(n-3)!$ poles picked up in total, thus providing a new type of expansion of its field theory conterpart into $(n-3)!$ terms, each of which comes as the zero slope limit of a single residue. This method also serves as a convenient way for numerical evaluation of string amplitudes, since one only has to deform the contours to get rid of pinching poles in order to resolve ill-defined contours, after which a direct expansion of the integrand with respect to $\alpha'$ is justified. There are several possible future explorations of interest.

Firstly, the Mellin-transformed formula takes Beta functions as the only building block. Since Beta functions are naturally associated to a $4$-pt scattering process, this decomposition of higher-multiplicity scattering into a product of Beta functions mimics the usual construction of Feynman diagrams out of a single interaction vertex (glued by propagators), or out of $3$-pt amplitudes by BCFW. This tends to suggest that the representation discussed in this paper may have a diagramatic interpretation, in which each individual Beta function in the integrand may carry certain physical or geometrical meaning. Actually, the representation of generlized Veneziano model in terms of Beta functions has already made an appearance in the very early days of string, e.g., \cite{Hopkinson:1969er}\footnote{The author would like to thank Stephan Stieberger for bringing this paper to the author's knowledge.}. Interestingly, the representation there also takes the form of an infinite series, which shares a very similar structure as the result we obtained in \eqref{infiniteseries} as we pushed the Mellin contour to infinity. Moreover, leading terms in the zero slope limit are also fully absorbed in the first term. While that representation is discussed in a special context, it is accompanied by a natural diagramatic interpretation. This is a strong indication that similar things may apply to the representation in this paper.

Secondly, notice the total counting of the number of poles relavent at leading order is $(n-3)!$, which arises from the well-organised structure among these poles. This number is particularly interesting since it has been repeated witnessed in our understanding of scattering amplitudes emerged in recent years inspired by Bern-Carrasco-Johansson (BCJ) relations~\citep{Bern:2008qj}. In particular, in a new type of formula based on scattering equations, for scattering of massless scalars in the doubly-colored $\phi^3$ theory~\citep{Cachazo:2013gna,Cachazo:2013iea}, the analog of disk integral is
\begin{equation}
m[\alpha|\beta]=\frac{1}{\text{vol SL}(2,\mathbb{C})} \int \frac{d^n\sigma_a}{\sigma_{\alpha(1),\alpha(2)}\cdots\sigma_{\alpha(n),\alpha(1)}}\frac{\prod'\delta(\sum_{b\neq a}\frac{s_{a,b}}{\sigma_{a,b}})}{\sigma_{\beta(1),\beta(2)}\cdots\sigma_{\beta(n),\beta(1)}},
\end{equation}
which produces all planar diagrams simultaneously consistent with the orderings of both Parke-Taylor factors in the integrand. In evaluating this formula one has to solve the scattering equations
\begin{equation}\label{scatteringequtaions}
\prod_{b\neq a}\frac{s_{a,b}}{\sigma_{a,b}}=0,
\end{equation}
which gives exactly $(n-3)!$ solutions for $\{\sigma_a\}$, each of which comes as the residue upon a pole that the prescribed contour encircles. Hence this also provides a natural $(n-3)!$ expansion of field theory amplitudes~\citep{Cachazo:2013gna}. In fact, this expansion has been shown to have a close connection with BCJ and KLT relations. Furthermore, $m[\alpha|\beta]$ itself is the inverse of the $(n-3)!\times(n-3)!$ based momentum kernal $\mathcal{S}[\alpha|\beta]$~\cite{Cachazo:2013iea}, and is exactly the zero slope limit of the disk integral \eqref{diskdef} as well~\citep{Broedel:2013tta,Cachazo:2013iea,Stieberger:2014hba}. Since Mellin transformations provides a different type of $(n-3)!$ expansion of field theory amplitudes inspired by the zero slope limit of string amplitudes, it would be interesting to see if there are any connections between the two.

\begin{acknowledgments}
EYY is grateful to Freddy Cachazo for collaboration at the early stages of this work, as well as Stephan Stieberger and Humberto Gomez for many useful comments on the draft. This work is supported by Perimeter Institute for Theoretical Physics. Research at Perimeter Institute is supported by the Government of Canada through Industry Canada and by the Province of Ontario through the Ministry of Research \& Innovation.
\end{acknowledgments}


\bibliographystyle{JHEP}
\bibliography{String_Mellin}

\end{document}